\documentclass[conference]{IEEEtran}
\pdfoutput=1

\def\BibTeX{{\rm B\kern-.05em{\sc i\kern-.025em b}\kern-.08em
    T\kern-.1667em\lower.7ex\hbox{E}\kern-.125emX}}

\usepackage{amsmath,amsfonts}
\usepackage{graphicx}

\usepackage{fancyhdr}
\usepackage{framed}
\usepackage{xcolor}

\colorlet{shadecolor}{gray!30}

\usepackage{csquotes}		
\usepackage{hyperref}
\usepackage{xspace}
\usepackage{subcaption}
\usepackage{booktabs}

\renewenvironment{quote}
  {\begin{shaded*}\small\list{}{\rightmargin=0.1cm \leftmargin=0.1cm}
   \item\relax}
  {\endlist\end{shaded*}}

\newcommand{\survey}{figures/}

\newcommand{\numParticipants}{102\xspace}%
\newcommand{\percentCommercialProjects}{84.3\,\%\xspace}%
\newcommand{\percentServer}{35.4\,\%\xspace}%
\newcommand{\percentDesktop}{27.4\,\%\xspace}%

\newcommand{\percentDeveloperDiscovered}{38\,\%\xspace}%
\newcommand{\percentCustomerDiscovered}{25\,\%\xspace}%
\newcommand{\percentTesterDiscovered}{25\,\%\xspace}%

\newcommand{\percentSemanticBugs}{69.6\,\%\xspace}
\newcommand{\percentMemoryBugs}{6.9\,\%\xspace}
\newcommand{\percentConcurrencyBugs}{8.8\,\%\xspace}

\newcommand{\percentBugsNotReproduced}{5\,\%\xspace}%

\newcommand{\percentEasyToReproduce}{71\,\%\xspace} %
\newcommand{\percentEasyToFix}{68\,\%\xspace} %
\newcommand{\percentEasyToLocalize}{39\,\%\xspace} %

\newcommand{\percentAnswerNumFiles}{87.3\,\%\xspace}%
\newcommand{\percentAnswerNumStatements}{56.9\,\%\xspace}%
\newcommand{\surveyNumberChangedStatementsAvg}{4.9\xspace}
\newcommand{\surveyNumberChangedStatementsMedian}{2\xspace}

\newcommand{\percentSingleLineFix}{36.2\,\%\xspace}

\begin{document}

\title{What we can learn from how programmers debug their code}

\makeatletter
\patchcmd{\@maketitle}
  {\addvspace{0.5\baselineskip}\egroup}
  {\addvspace{-1\baselineskip}\egroup}
  {}
  {}
\makeatother

\author{\IEEEauthorblockN{Thomas Hirsch}
\IEEEauthorblockA{\textit{Institute of Software Technology}\\
\textit{Graz University of Technology}\\
Graz, Austria\\
thirsch@ist.tugraz.at}
\and
\IEEEauthorblockN{Birgit Hofer}
\IEEEauthorblockA{\textit{Institute of Software Technology}\\
\textit{Graz University of Technology}\\
Graz, Austria\\
bhofer@ist.tugraz.at}
}

\maketitle

\begin{abstract}
Researchers have developed numerous debugging approaches to help programmers in the debugging process, 
but these approaches are rarely used in practice. %
In this paper, we investigate how programmers debug their code %
and what researchers should consider when developing debugging approaches. We conducted an online questionnaire where \numParticipants~programmers provided information about recently fixed bugs. We found that the majority of bugs (\percentSemanticBugs) are semantic bugs. Memory and concurrency bugs do not occur as frequently (\percentMemoryBugs and \percentConcurrencyBugs), but they consume more debugging time. %
Locating a bug is more difficult than reproducing and fixing it.
Programmers often use only IDE build-in tools for debugging. 
Furthermore, programmers frequently use a replication-observation-deduction pattern when debugging.
These results suggest that debugging support is particularly valuable for memory and concurrency bugs. Furthermore, researchers should focus on the fault localization phase and integrate their tools into commonly used IDEs.
\end{abstract}

\begin{IEEEkeywords}
debugging in practice, user questionnaire
\end{IEEEkeywords}

\thispagestyle{fancyplain}
\footnotetext{\textcopyright 2021 IEEE.  Personal use of this material is permitted.  Permission from IEEE must be obtained for all other uses, in any current or future media, including reprinting/republishing this material for advertising or promotional purposes, creating new collective works, for resale or redistribution to servers or lists, or reuse of any copyrighted component of this work in other works.}

\section{Introduction}

Debugging consumes about 30-90\,\% of the total development time~\cite{Ang2017}.
Researchers have proposed numerous debugging techniques \cite{WongSurvey2016,Gazzola2019} to support developers in this task, but these techniques are rarely used in practice. 
One reason %
might be that academic  approaches do not satisfy the requirements that arise when developing software in practice.

Several researchers have conducted surveys with programmers to gain insights on how programmers debug their code.
Eisenstadt~\cite{Eisenstadt1997} conducted an online survey with professional software developers where they described their most demanding bugs.
Siegmund \textit{et al.}~\cite{Siegmund2014} observed eight programmers in their debugging process.
All of them executed a simplified scientific method by building and checking hypotheses.
Perscheid \textit{et al.}~\cite{Perscheid2017} conducted an online questionnaire with 300 participants. %
The majority of the participants indicated to regularly use print statements. %

We continue this line of research by conducting an online questionnaire where we ask professional developers questions about a recently fixed bug. %
The questions comprise the bug's impact and root cause, amount of changes, debugging time, %
and perceived difficulty.
While \cite{Eisenstadt1997} and \cite{Perscheid2017} asked programmers to describe their most difficult bug, we asked them to describe their most recent bug fix.
Therefore, the results of this study represent typical faults instead the most difficult bugs.

\begin{figure*}[t]
    \centering
        \begin{subfigure}[b]{0.32\textwidth}
        \includegraphics[width=\textwidth,trim={0.3cm 0.66cm 0.3cm 0.7cm},clip]{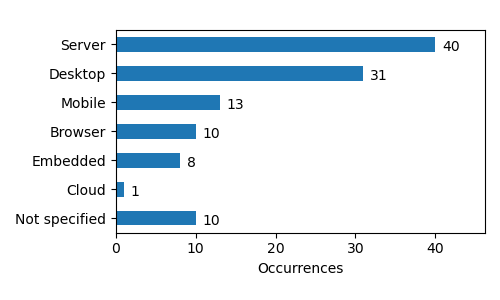}
        \caption{Application type}
        \label{fig:platform}
    \end{subfigure}
		\, 
    \begin{subfigure}[b]{0.32\textwidth}
        \includegraphics[width=\textwidth,trim={0.3cm 0.66cm 0.28cm 0.7cm},clip]{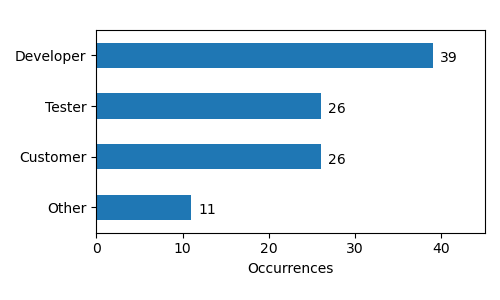}
        \caption{Who discovered the bug}
        \label{fig:bugDiscoverer}
    \end{subfigure}
		\begin{subfigure}[b]{0.32\textwidth}              %
        \includegraphics[width=\textwidth,trim={0.3cm 0.66cm 0.28cm 0.7cm},clip]{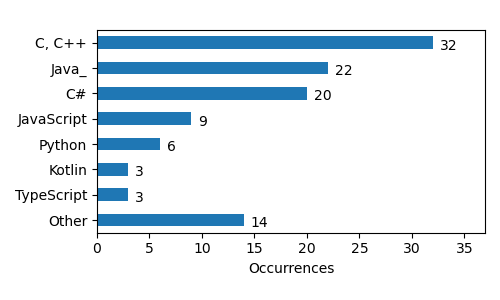}
        \caption{Used languages}
        \label{fig:language}
    \end{subfigure}
    \caption{General results}\label{fig:questionnaire}
\end{figure*}
\section{Study Set-up}\label{sec:setup}

\emph{Objectives:}
To bring research in the field of automatic debugging closer to practice, we need to know what types of bugs are most common and how much time it takes to fix them.
This questionnaire collects data on failure and fault types, the time required to reproduce, localize, and correct a bug, the amount of necessary code changes, and the used debugging method.
The goal is to collect information about typical bugs, not the most difficult bugs.
Therefore, we ask developers to fill out the questionnaire for their most recent bug fix.

\emph{Target population:}
This survey addresses developers who fix bugs. %
It does not address testers, product owners, or managers.

\emph{Bug types:}
We used the  Root cause and Impact dimension of 
Tan \textit{et al.} \cite{Tan2014a} as bug types.
The \textit{Impact} dimension describes how a bug has manifested.
The original version~\cite{Tan2014a} comprises six categories:
\begin{itemize}
	\item \textit{Hang}: The software does not respond.
	\item \textit{Crash}:	  The software halts abnormally.
	\item \textit{Data corruption}:	 User data is mistakenly modified.
	\item \textit{Performance degradation}:	 The software works slowly.
	\item \textit{Incorrect functionality}:	 The software behaves incorrectly.
	\item \textit{Other}:	 This includes all other impacts. %
\end{itemize}
Since we wanted to keep the number of entries in the \textit{Other} category as small as possible, we added two more categories:
\begin{itemize}
	\item \textit{No Build}~\cite{Chillarege92}:	 The software is not buildable.
	\item \textit{Security}~\cite{IEEE2010}:	 There is a security vulnerability.
\end{itemize}
The \textit{Root cause} dimension~\cite{Tan2014a} describes the cause of the bug:
\begin{itemize}
	\item \textit{Semantic}:  Bugs due to inconsistency of requirements / programmers intentions and actual implementation, e.g. missing corner case, incorrect algorithm, incorrect flow.
	\item \textit{Memory}:	 Bugs arising from improper/incorrect memory handling, e.g. memory leaks, buffer overflows.
	\item \textit{Concurrency}:  Bugs arising from improper synchronization and other concurrency issues, e.g. deadlocks, race conditions.
\end{itemize}
We added the category \textit{Other} to allow programmers to indicate faults in the documentation, and the build system:
\begin{itemize}
	\item \textit{Other}:	 All other bugs, e.g. documentation errors, issues in build scripts, bugs due to misconfigurations.
\end{itemize}

\emph{Methodology:}
We created Google forms with questions in German and English. 
We asked the programmers to fill out the questionnaire immediately after they have fixed a bug.
The questions focus  on a specific bug rather than on debugging in general and cover
 the type of the project %
 and the programming language, 
 the impact and root cause, %
 the time it took to reproduce, locate, and fix the bug,
 the perceived difficulty,
 the amount of code changes, and
 the used debugging approach.
The online appendix contains the complete list of questions. %

To make sure that the questions are unambiguous and that participants are able to answer them in a reasonable amount of time, we conducted a pilot study in form of a thinking-aloud test with nine software developers. %
In this test, the test participants were observed when filling out the survey and they were encouraged by the observer to communicate when they were unsure about the meaning of a question.
Furthermore, the test participants explained the bug to the observer before filling out the survey so that the observer could check whether the test participants were able to correctly classify their bugs.
All participants of the pilot study were able to answer the questions in less than five minutes.

The questionnaire was  advertised on the project website and on LinkedIn (both via posts and personal messages).
After the completion of the questionnaire, the participants were encouraged to fill out the survey for another bug and to share it with their colleagues.
The participation was anonymous, voluntary, and no compensation was paid.
The questionnaire was open from March to May~2020 and we received \numParticipants~answers.

\section{Study results}\label{sec:results}

\emph{General information.}
\percentCommercialProjects of the bugs are from commercial software projects; the other bugs are from open-source projects.
The majority of the projects were server (\percentServer) and desktop (\percentDesktop) applications (Fig.~\ref{fig:platform}).
\percentDeveloperDiscovered of the bugs were discovered by the developers themselves,
\percentCustomerDiscovered were discovered by customers and \percentTesterDiscovered were discovered by testers (Fig.~\ref{fig:bugDiscoverer}).
The most used languages are C, C++, Java, and C\# (Fig.~\ref{fig:language}).
The \enquote*{Other} category includes programming languages which were mentioned only once or twice. %

\emph{Root cause and impact.}
The reported bugs are quite diverse ranging from \enquote*{time zone issues} and \enquote*{wrong units} over \enquote*{data validation too strict} to \enquote*{outdated configurations}.
Table~\ref{tab:RootCausesAndImpacts} shows the occurrences of the different root causes and impacts.
The majority of the bugs manifest themselves as incorrect functionality.
Semantic bugs are the major root cause.

\begin{table}[htbp]
	\centering
	\caption{Number of root causes and impacts}
	\label{tab:RootCausesAndImpacts}
	\small
			\begin{tabular}{lrrrr|r}
\toprule
\textbf{Impact/Root cause} &  \textbf{Sem.} &  \textbf{Mem.} &  \textbf{Conc.} &  \textbf{Other} &  \textbf{Total} \\
\midrule
Crash                   &        10 &       6 &            0 &      1 &     17 \\
Data corruption         &         2 &       0 &            1 &      0 &      3 \\
Hang                    &         1 &       0 &            3 &      1 &      5 \\
Incorrect build         &         0 &       0 &            0 &      3 &      3 \\
Incorrect functional. &        50 &       1 &            3 &      7 &     \textbf{61} \\
Performance degrad. &         3 &       0 &            0 &      0 &      3 \\
Security                &         2 &       0 &            0 &      2 &      4 \\
Other                   &         3 &       0 &            2 &      1 &      6 \\
\midrule \textbf{Total}                   &        \textbf{71} &       7 &            9 &     15 &    102 \\
\bottomrule
\end{tabular}
\end{table}

\begin{table}[htbp]
	\caption{Number of files, methods, and statements changed. \enquote*{-} indicates the number of participants who did not answer.}%
	\label{tab:NumberOfFilesMethodsAndStatementsChanged}
	\centering
		\small
		\begin{tabular}{lrrrrrr}
\toprule
\textbf{Changes}&		\textbf{-}	& \textbf{0}	& \textbf{1 (\% )}	&\textbf{[2-5]} &	\textbf{[6-10]} & \textbf{$>$10} \\ \midrule
\textbf{Files} \\
 Semantic    & 9 & 0 & 40 (64.5\,\%)& 21 & 0 & 1 \\
 Memory      & 1 & 0 & 5  (83.3\,\%)& 0 & 0 & 1 \\
 Concurrency & 1 & 0 & 7  (87.5\,\%)& 1 & 0 & 0 \\
 Other       & 2 & 0 & 8  (61.5\,\%)& 2 & 1 & 2 \\ \cline{2-7}
 Total       & 13 & 0 & 60 (\textbf{67.4\,\%})& 24 & 1 & 4 \\

\midrule
\textbf{Methods} \\ 
 Semantic     & 12& 0 & 32 (54.2\,\%)& 22 & 3 & 2 \\
 Memory       & 1 & 0 & 5 (83.3\,\%) & 0 & 0 & 1 \\
 Concurrency  & 2 & 0 & 4 (57.1\,\%) & 3 & 0 & 0 \\
 Other        & 4 & 3 & 5 (45.5\,\%) & 1 & 2 & 0 \\ \cline{2-7}
 Total        & 19& 3 & 46 (\textbf{55.4\,\%}) & 26 & 5 & 3 \\
\midrule
\textbf{Statements} \\
 Semantic     & 29& 0 & 12 (28.6\,\%) & 20 & 3 & 7 \\
 Memory       & 2 & 0 & 4 (80.0\,\%) & 1 & 0 & 0 \\
 Concurrency  & 4 & 0 & 2 (40.0\,\%) & 3 & 0 & 0 \\
 Other        & 9 & 0 & 3 (50.0\,\%) & 2 & 0 & 1 \\ \cline{2-7}
 Total        & 44& 0 & 21 (\textbf{36.2\,\%}) & 26 & 3 & 8 \\
\bottomrule
		\end{tabular}%
\end{table}

\emph{Amount of changes.}
Table~\ref{tab:NumberOfFilesMethodsAndStatementsChanged} provides an overview of the number of changed files, methods, and statements to fix the bug.
Not all study participants provided information about the amount of code changes:
\percentAnswerNumFiles indicated the number of files, but only \percentAnswerNumStatements indicated the number of statements they had changed. %
The three answers with zero changed methods for the Other category can be explained with bugs in the documentation or build system where actually no methods were changed.
67.4\,\% of all bugs were corrected by making changes in a single file,
55.4\,\% by making changes in only one method, and
only 36.2\,\% of the bugs were corrected by changing a single statement.

\emph{Time analysis.}
Table~\ref{tab:MedianTimeInHours} shows the median and mean time required for reproducing, locating, and fixing the bugs for all root causes and selected impacts.
We do not provide this information for the other impact categories because of the low number of occurrences.
In general, locating faults is more time-consuming than fixing faults.
Memory and concurrency bugs take the longest to  reproduce and locate.
The high mean values for memory and concurrency bugs compared to their medians indicate that there is a high variation in the time required to locate and fix such bugs.
\percentBugsNotReproduced of the bugs were not reproduced because either the bug was obvious from the log-file or stack trace or it was impossible to reproduce it. %

\begin{table}[tbp]
	\centering
		\caption{Time to reproduce, locate, and fix faults (in hours)}
	\label{tab:MedianTimeInHours}
	\small
				\begin{tabular}{l|rr|rr|rr}
				\toprule
				& \multicolumn{2}{c|}{\textbf{Reproduce}} & \multicolumn{2}{c|}{\textbf{Locate}} & \multicolumn{2}{c}{\textbf{Fix}} \\
				{} &  mean &  med. &  mean & med. &  mean & med. \\
				\midrule

\textbf{Root cause} &&&&&& \\ 
Concurrency &            \textbf{10.3} &               2.0 &         \textbf{11.7} &            4.0 &       \textbf{7.3} &         1.0 \\
Memory      &            \textbf{19.1} &               5.0 &         \textbf{24.8} &            4.0 &       1.3 &         1.0 \\
Semantic    &             1.0 &               0.5 &          2.2 &            1.0 &       1.5 &         1.0 \\
Other       &             0.6 &               0.1 &          1.4 &            0.5 &       2.4 &         1.0 \\

				\midrule

\textbf{Impact}             &                 &                   &              &                &           &             \\

Crash                   &             8.4 &               0.5 &         10.2 &            1.0 &       2.1 &         1.0 \\
Hang                    &             3.8 &               2.5 &          6.3 &            1.0 &       1.2 &         1.0 \\
Incorrect funct. &             0.9 &               0.5 &          2.2 &            0.7 &       1.3 &         0.8 \\

				\midrule
\textbf{Total}       &             3.0 &               0.5 &          4.4 &            1.0 &       2.1 &         1.0 \\
				\bottomrule
				\end{tabular}
\end{table}

\emph{Debugging tools and methods.}
Figure~\ref{fig:debuggingMethods} shows which debugging methods have been used to localize the bugs.
Breakpoints, step-by-step execution, and print/log statements are the most-used debugging methods.
The \enquote*{Others} category includes optional free text responses, where
participants mentioned (unit) tests, profilers, additional scripting, code inspection, and crash reports.

\begin{figure}
	\centering
		\includegraphics[width=0.8\columnwidth,trim={0.2cm 0.6cm 0.3cm 0.55cm},clip]{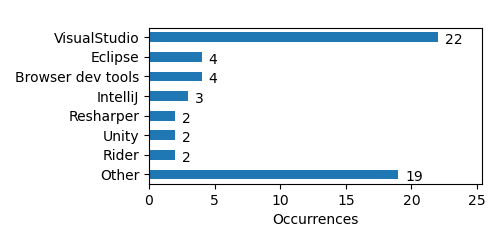}
		\caption{Debugging tools}
        \label{fig:debuggingTools}
\end{figure}

The majority of the study participants indicated that they used only the IDE's built-in debugging tools for support (Figure~\ref{fig:debuggingTools}).
Visual Studio was mentioned most-often by far, followed by Eclipse, browser development tools (e.g. for Chrome), and IntelliJ.
The \textit{Other} category summarizes all tools that were mentioned only once.

We asked the participants to describe in their own words how they proceeded to localize the bugs.
From their answers, we have identified a replication-observation-deduction pattern in the majority of the responses, as the following examples demonstrate:

\begin{figure}
	\centering
        \includegraphics[width=0.8\columnwidth,trim={0.25cm 0.5cm 0.3cm 0.7cm},clip]{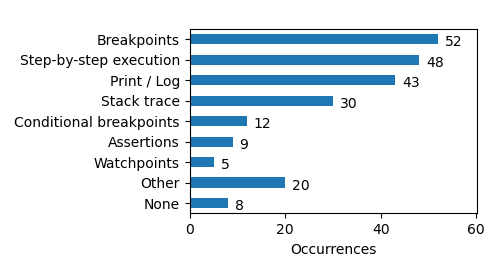}
        \caption{Debugging methods}
        \label{fig:debuggingMethods}
\end{figure}

\begin{quote}reproduce bug [\textbf{replication}], set breakpoints, deeper and deeper debugging with a colleague [\textbf{observation/deduction}]
\end{quote}

\begin{quote}Deployed a local instance of the project to replicate the issue [\textbf{replication}]. Set a debug point in the area of the code that handles the given logic, checked the outcome of the if statement  [\textbf{observation}] and located the problem in how the logical operators were being used [\textbf{deduction}].
\end{quote}

In some cases, participants omitted trivial steps:
\begin{quote}logical thinking, exclusion process [\textbf{deduction}], advanced logging [\textbf{observation}]
\end{quote}
\begin{quote}Follow failure description provided by user to reproduce the bug [\textbf{replication}]. Debug data flow from frontend to backend and identify data inconsistency between those [\textbf{observation}].
\end{quote}

A small number of participants did not use the replication-observation-deduction approach.
Instead, they only analyzed the code.
One participant mentioned \enquote{backtracking} and two participants indicated to use \enquote{code inspection} techniques. %

\begin{figure*}
    \centering
        \includegraphics[width=\textwidth,trim={0.3cm 0.6cm 0.3cm 0.7cm},clip]{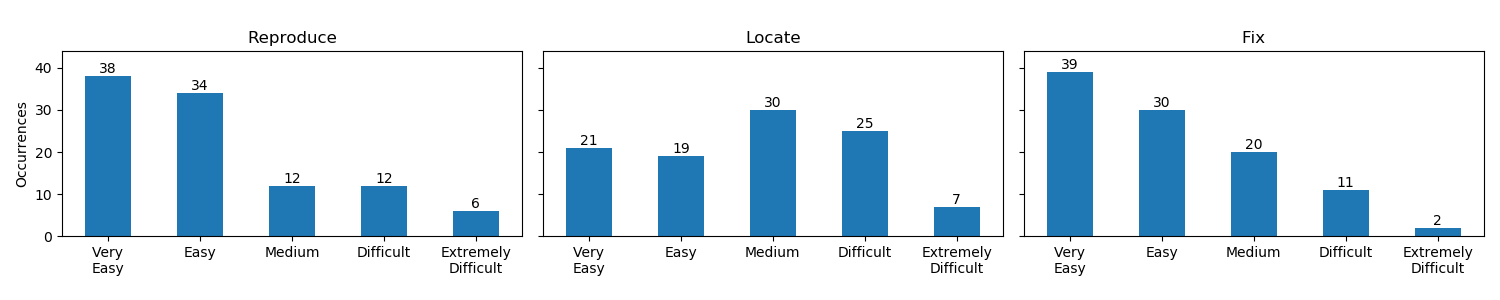}
    \caption{Perceived difficulty to reproduce, locate, and fix bugs}
		\label{fig:perceivedDifficulty}
\end{figure*}

\emph{Perceived Difficulty.}
Figure~\ref{fig:perceivedDifficulty} shows the perceived difficulty for reproducing, localizing, and fixing the faults.
While the majority of the participants indicated that it was \enquote*{easy} or \enquote*{very easy} to reproduce (\percentEasyToReproduce) and fix (\percentEasyToFix) their bug, only \percentEasyToLocalize indicated the same for finding the faulty source code location.

\section{Threats to validity}\label{sec:threats}

The selection of the study participants is a threat to the internal validity of the study.
We received the majority of the responses after writing personal messages to former fellow students.
Therefore, there is a strong bias w.r.t. the educational background of the study participants and their programming experience.
Furthermore, the study participants might not answer the questionnaire on a recently fixed bug, but on an extremely difficult bug. %
The number of changed statements in the survey is low (average; \surveyNumberChangedStatementsAvg,  median: \surveyNumberChangedStatementsMedian).
Developers might not have provided this number for bugs with many changes, because they did not want to count them.
Moreover,  participants might over- or underestimate the time they actually needed to reproduce, localize, and fix the bug.
Some participants might misunderstand some questions or wrongly categorize their bug.
We have addressed the last threat by performing a thinking-aloud test with nine software developers which helped us to reveal problems.

\section{Conclusion and future work}\label{sec:conclusion}
We conducted an online questionnaire with \numParticipants programmers were we asked questions about a recently fixed fault.
From the answers of this questionnaire, we draw the following conclusions:

\emph{Single faulty line assumption.}
The single faulty line assumption is unrealistic.
Only \percentSingleLineFix of the survey participants indicated that they fixed their bugs by changing a single statement. %
This confirms the findings of Lucia \textit{et al.}~\cite{Lucia2012}: %
They analyzed the fixes of three software projects and found that  less than 40\,\% of the bugs were fixed by changing a single line. %
Researchers have to take this into account when designing and evaluating new debugging tools.
It is not sufficient to highlight one suspicious line at a time because faults are usually more complex.

\emph{Focus on fault localization phase.}
Locating faults is more time-consuming than fixing them.
This implies that debugging support is particularly valuable for the fault localization process.

\emph{Focus on certain types of bugs.}
While semantic faults are the most frequent root causes, developers spent more time on reproducing and locating memory and concurrency bugs.
Therefore, debugging tools for these fault types could offer
significant time savings.

\emph{Support of the natural reasoning process.}
Programmers follow a replication-observation-deduction pattern when debugging because this pattern allows them to build a mental model of the program flow.
A debugging tool supporting this natural reasoning process might be more accepted by programmers. 

\emph{Integration into widely used IDEs.}
Developers use their IDEs when debugging instead of stand-alone debugging tools.
Therefore, debugging tools have to be  integrated in IDEs to increase acceptance and enable wide spread use.

In future work, we will recruit more programmers to fill out the questionnaire.
Furthermore, 
we will compare the results of this survey with the status quo in an open-source project.
While a survey offers a broad overview over different programming languages, domains, and applications, the investigation of a single open-source project will provide a holistic picture of one project. %

\section{Data Availability}
The catalogue of questions and all answers are publicly available at \url{https://doi.org/10.5281/zenodo.4449045}.

\section*{Acknowledgment}
The work described in this paper has been funded by the Austrian Science Fund (FWF): P 32653 (Automated Debugging in Use).

\bibliographystyle{IEEEtran}
\bibliography{library}

\end{document}